\documentclass[12pt]{article}
\usepackage{epsfig,cite}
\usepackage{amsmath}
\usepackage{amssymb}
\usepackage{amsfonts}
\usepackage{latexsym}
\usepackage{wasysym}
\usepackage{bm}
\catcode`@=11 %This allows us to modify plain macros
\def\slash#1{\mathord{\mathpalette\c@ncel#1}}
 \def\c@ncel#1#2{\ooalign{$\hfil#1\mkern1mu/\hfil$\crcr$#1#2$}}
\def\lsim{\mathrel{\mathpalette\@versim<}}
\def\gsim{\mathrel{\mathpalette\@versim>}}
 \def\@versim#1#2{\lower0.2ex\vbox{\baselineskip\z@skip\lineskip\z@skip
       \lineskiplimit\z@\ialign{$\m@th#1\hfil##$\crcr#2\crcr\sim\crcr}}}

\def\ytop{y_t}
\def\bea{\begin{eqnarray}}
\def\eea{\end{eqnarray}}
\def\beq{\begin{equation}}
\def\eeq{\end{equation}}
\catcode`@=11 %This allows us to modify plain macros
\def\slash#1{\mathord{\mathpalette\c@ncel#1}}
 \def\c@ncel#1#2{\ooalign{$\hfil#1\mkern1mu/\hfil$\crcr$#1#2$}}
\def\lsim{\mathrel{\mathpalette\@versim<}}
\def\gsim{\mathrel{\mathpalette\@versim>}}
 \def\@versim#1#2{\lower0.2ex\vbox{\baselineskip\z@skip\lineskip\z@skip
       \lineskiplimit\z@\ialign{$\m@th#1\hfil##$\crcr#2\crcr\sim\crcr}}}
\catcode`@=12 %at signs are no longer letters
\def\twiddles#1{\mathrel{\mathop{\sim}\limits_
                        {\scriptscriptstyle {#1}}}}

\def\mtop{m_t}
\def\mhiggs{m_H}
\def\({\left(}
\def\){\right)}
\def\[{\left[}
\def\]{\right]}

\def\as{\alpha_s}

\relax

\def    \hepph  #1 {{\tt hep-ph/#1}}
\def    \hepex  #1 {{\tt hep-ex/#1}}
\def\shat{\hat s}

\newsavebox\tmpfig

\begin{document}

\pagestyle{empty}

\begin{flushright}

IFUM-911-FT\\ Edinburgh 2008/1\\CERN-PH-TH/2008-009
\end{flushright}

\begin{center}
\vspace*{0.5cm}
{\Large \bf Higgs production via gluon-gluon fusion\\ with finite top
  mass beyond next-to-leading order} 
 \\
\vspace*{1.5cm}
Simone Marzani,$^{a}$, Richard D.~Ball,$^{a,b}$ Vittorio Del
Duca,$^{c}$\footnote{On leave from INFN, Sezione di Torino, Italy} \\
Stefano~Forte$^{d}$ and Alessandro Vicini$^{d}$
\\
\vspace{0.3cm}  {\it
{}$^a$School of Physics, University of Edinburgh,\\
Edinburgh EH9 3JZ, Scotland, UK\\ \medskip
{}$^b$CERN, Physics Department, Theory Division,\\
CH-1211 Gen\`eve 23, Switzerland\\\medskip
{}$^c$INFN, Laboratori Nazionali di Frascati,\\
Via E.~Fermi 40, I-00044 Frascati, Italy\\\medskip
{}$^d$Dipartimento di Fisica, Universit\`a di Milano and
INFN, Sezione di Milano,\\
Via Celoria 16, I-20133 Milano, Italy}\\
\vspace*{1.5cm}

{\bf Abstract}
\end{center}

\noindent
We present a computation of 
the cross section for inclusive Higgs
production in gluon--gluon fusion for finite values of the top mass
in perturbative QCD to all orders in the limit of high partonic
center--of--mass energy. 
We show that at NLO the high energy contribution accounts for most of the
difference between  the result found with finite top mass and that
obtained in the limit $\mtop\to\infty$. We use our result to improve
the known NNLO order result obtained at
$\mtop\to\infty$.
We estimate the effect of the high energy NNLO $m_t$ dependence on the 
$K$ factor to be of the order of a few per cent.

\vspace*{1cm}

\vfill
\noindent

\begin{flushleft} CERN-PH-TH/2008-009\\
January 2008 \end{flushleft}
\eject

\setcounter{page}{1} \pagestyle{plain}
\section{The cross section in the soft limit and in the  hard limit}

The determination of higher--order corrections to collider
processes~\cite{salam}, and specifically Higgs production~\cite{harlander}
in perturbative QCD is becoming increasingly important in view of
forthcoming phenomenology at the LHC.
The dominant Higgs production mechanism in the standard model is
inclusive gluon--gluon fusion ($gg\to H+X$) through a top loop. The next--to--leading order
corrections to this process were computed several years
ago~\cite{graudenz,spira}  and turn out to be very large (of order
100\%). The bulk of this large correction comes from the radiation of soft
and collinear gluons~\cite{kramer}, which give the leading
contribution in the soft limit in which the partonic center-of-mass
energy $\shat$ tends to the Higgs mass $m_H^2$, and which at LHC energies turns out to
dominate the hadronic cross section after convolution with the parton
distributions.   

This dominant contribution
does not resolve the
effective gluon-gluon-higgs (ggH) coupling induced by the top loop. As a
consequence, the NLO correction can be calculated~\cite{djouadi,dawson} quite accurately in the limit
$\mtop\to\infty$, where it simplifies considerably because the ggH
coupling becomes pointlike and  the corresponding Feynman diagrams
have one less loop. Recently, the NNLO corrections to this process
have been computed in the $\mtop\to\infty$ limit~\cite{anastasiou}. The 
NNLO result appears to be perturbatively quite stable, and this
stability is confirmed upon inclusion~\cite{moch} of terms in the next few orders
which are logarithmically enhanced as $\shat\to\mhiggs$, which can be
determined~\cite{deflorian} using soft resummation methods. This
suggests that also at NNLO the large $\mtop$ approximation should
provide a good approximation to the yet unknown exact
result. 

However, this is only true for the total inclusive cross section: for
example, if one looks at the production of Higgs plus jets, if the
transverse momentum is large the infinite $\mtop$ approximation
fails~\cite{delduca}. Indeed,  even 
though the $\mtop$--independent contribution from soft and
collinear radiation turns out to dominate the cross section at the
hadronic level, it does not necessarily provide a good approximation
to the partonic cross section in a fixed kinematical region.
 In
particular, the infinite $\mtop$ approximation, which becomes exact in the
soft limit,  fails in the opposite (hard)
limit of large center--of--mass energy. This is due to the fact that
the ggH vertex is pointlike in the infinite $\mtop$ limit, whereas for
finite $\mtop$  the quark loop provides a form factor
(as we shall see explicitly below). Clearly,
a point--like
interaction has a completely different high energy
behaviour than a resolved interaction which is softened by a form
factor: in fact one can show~\cite{catani} that 
a point--like interaction at
$n$--th perturbative order has double energy logs while a resolved
interaction has only single logs. 

This means that as $\shat\to\infty$ the 
$gg\to H+X$ partonic cross section $\hat \sigma$ behaves as 
\beq
\label{hebeh}
\hat\sigma\twiddles{\shat\to\infty}\left\{\begin{array}{c}\sum_{k=1}^\infty\as^k\ln^{2k-1}\left(\frac{\shat}
{\mhiggs^2}\right)\qquad\hbox{pointlike: }\mtop\to\infty\\  \\
\sum_{k=1}^\infty\as^k\ln^{k-1}\left(\frac{\shat}
{\mhiggs^2}\right)\qquad\hbox{resolved: finite }\mtop\\
\end{array} \right. 
\eeq
Hence, as the center-of-mass energy grows, eventually $\mtop\to\infty$
ceases to be a good approximation to the exact result. It is clear
from eq.~(\ref{hebeh}) that this high energy  deviation  between the
exact and approximate behaviour is stronger at higher orders, so one
might expect the relative accuracy of the infinite $\mtop$
approximation to the $k$--th order perturbative contribution to the
cross section to become worse as the perturbative order increases. 
Conversely, this suggests that it might be worth determining the
high energy behaviour of the exact cross section and use the result  to
improve the infinite $\mtop$ result, which is much less difficult to determine.
Eventually, a full resummation of these contributions might also
become relevant.

The 
leading high energy contributions to this process in the infinite
$\mtop$ limit have in fact been computed some time ago in
Ref.~\cite{hautmann}: this amounts to a determination of the coefficient of
the double logs eq.~(\ref{hebeh}), in the pointlike case. In this paper, we
compute the coefficients of the single logs eq.~(\ref{hebeh}) in the
resolved (exact) case. Our result takes the form of a double integral,
whose numerical evaluation order by order in a Taylor expansion
gives the coefficient of the
logs eq.~(\ref{hebeh}) (at the lowest perturbative
order the integral can be computed in closed form). After 
checking our result against the known full NLO result of
Ref.~\cite{graudenz,spira}, we will discuss the way knowledge of the
exact high energy behaviour of the cross section at a given order
can be used to
improve the infinite $\mtop$ result, using the NLO case,
where everything is known, as a testing ground. We will show that in
fact, at NLO the different high energy behaviour eq.~(\ref{hebeh})
accounts for most of the difference between the exact and infinite
$\mtop$ cross sections. We will then repeat this analysis in the NNLO
case, where only the infinite $\mtop$ result is currently known. We
will show that in fact at this order the contribution of the
logarithmically enhanced terms which dominate the partonic
cross section at
high energy is substantial even for moderate values of the partonic
center-of-mass energy,
such as $\hat s\sim 2 \mhiggs^2$.

The calculation of the leading high energy logs is presented in
section~2, while in section~3 we discuss its use to improve the NLO and
NNLO results. The appendix collects the explicit expressions of the
form factors which parametrize the amplitude for the process $gg\to H$
with two off--shell gluons, which is required for the calculation of sect.~2.

\section{Determination of the leading high energy logarithms}
\subsection{Definitions, kinematics and computational procedure}
We compute the total inclusive partonic cross section
$\hat\sigma(gg\to H+X)$   in
an expansion in power of $\as$, as a function of
the partonic center-of-mass 
energy $\hat s$:
\beq\label{partxsec}
\hat\sigma(gg\to H+X)=\hat
\sigma_{gg}\left(\as;\tau;\ytop,\mhiggs^2\right),
\eeq
where the dimensionless variables $\tau$ and $\ytop$ parametrize
respectively the
partonic center-of-mass energy and the dependence on
the top mass:
\bea
\label{ydef}
\tau&\equiv& \frac{\mhiggs^2}{\shat
}\\
\label{zetdef}
\ytop&\equiv& \frac{\mtop^2}{\mhiggs^2}.
\eea
The corresponding contribution to the hadronic cross section
$\sigma$ can be obtained by convolution with the gluon-gluon parton
luminosity ${\cal L}$:
\bea\label{partohadr}
\sigma_{gg}(\tau_h;\ytop,\mhiggs^2)&=&\int_{\tau_h}^1\!dw\,  \hat \sigma_{gg}
\left(\as;\frac{\tau_h}{w};\ytop,\mhiggs^2\right){\cal L}(w)\\
\label{lumidef}
 {\cal L}(w)&\equiv&\int_w^1\!\frac{dx_2}{x_2}\,
 g_{h_1}\left(\frac{w}{x_2},\mhiggs^2\right)  g_{h_2}\left(x_2,\mhiggs^2\right),
\eea
where $g_{h_i}(x_i,Q^2)$  is the gluon distribution in the $i$-th
incoming hadron and in eq.~(\ref{partohadr}) 
the dimensionless variables $\tau_h$ parametrizes the hadronic
center-of-mass energy  $s$
\beq
\label{xdef}
\tau_h\equiv \frac{\mhiggs^2}{s}.
\eeq
Note that $0\le \tau_h\le \tau\le 1$, and that if  $\ytop < \frac{1}{4}$ then
the intermediate $t\bar t$ pair produced by the gluon-gluon fusion can
go on shell.

It is convenient to define a dimensionless hard coefficient function
$C(\as(\mhiggs^2);\tau,\ytop)$ 
\bea\label{cfdef}
\hat \sigma_{gg}
\left(\as;\tau;\ytop,\mhiggs^2\right)&=& \sigma_0(\ytop)
C(\as(\mhiggs^2),\tau,\ytop)\\
\label{cfexp}
C(\as(\mhiggs^2),\tau,\ytop)&=&\delta(1-\tau)+\frac{
  \as(\mhiggs^2)}{\pi}C^{(1)}(\tau,\ytop)+\left(\frac{
    \as(\mhiggs^2)}{\pi}\right)^2
C^{(2)}(\tau,\ytop),
\eea
where $\sigma_0\delta(1-\tau)$ is the leading order cross section,
determined long ago in ref.~\cite{higgslo}:
\beq\label{sigzero}
\sigma_0(\ytop)=\frac{\alpha_s^2 G_F \sqrt{2}}{256
  \pi}  \left|4 \ytop \left(1- \frac{1}{4} (1-4 \ytop) s^2_0(\ytop)\right)\right|^2\,,
\eeq
where 
\beq\label{szerodef}
 s_0(\ytop)=\left\{\begin{matrix}&\ln \left(\frac{1-\sqrt{1-4 \ytop}}{1+\sqrt{1-4
       \ytop}}\right)+ \pi\,i & \textrm{if} \quad \ytop<
 \frac{1}{4} \\
& 2\, i \tan^{-1} \left(\sqrt{\frac{1}{4 \ytop-1}}\,\right)=2\, i
\sin^{-1} \left(\sqrt{\frac{1}{4 \ytop}}\,\right) &\textrm{if} \quad
\ytop \ge \frac{1}{4}.\end{matrix}\right. 
\eeq
We also define the Mellin transform
\beq\label{cmeldef}
C(\as(\mhiggs^2),N,\ytop)=\int_0^1\!d\tau\, \tau^{N-1}C(\as(\mhiggs^2),\tau,\ytop),
\eeq
denoted with the same symbol by slight abuse of notation.

We are interested in the determination of the leading high energy
contributions to the partonic cross
section $\hat\sigma(gg\to
H+X)$, namely, the leading contributions to
$C(\as(\mhiggs^2),\tau,\ytop)$
as $\tau\to0$ to all orders in
$\as(\mhiggs^2)$. Order by order in $\as(\mhiggs^2)$, these 
correspond to the highest rightmost pole in
$N$ in the expansion in powers of $\as(\mhiggs^2)$ 
of $C(\as(\mhiggs^2),N,\ytop)$. The leading
singular contributions to the partonic cross section $\hat\sigma(gg\to
H+X)$ to all orders can be
extracted~\cite{catani} from the computation of the cross
section for a slightly  different process, 
namely, the cross section $\sigma_{\rm off}(gg\to
H)$ computed at leading order, but with incoming off-shell gluons,
a suitable choice of kinematics and a suitable prescription for the sum over
polarizations. 

The
procedure  used for this determination is based on the so-called
high energy (or $k_t$) factorization~\cite{catani},  and consists of
the following steps.

\begin{itemize}
\item 
One computes the  matrix element ${\cal M}_{ab}^{\mu\nu}(k_1,k_2)$ for the
leading-order process $gg\to
H$ at leading order with
two incoming off-shell gluons with polarization indices $\mu,\nu$ and
color indices $a,b$. The  momenta $k_1$, $k_2$ of the gluons
in the center-of-mass frame of the hadronic collision admit the
Sudakov decomposition at high energy
\beq\label{sudakov}
k_i=z_i p_i+{\bf k}_i,
\eeq
where $p_i$ are lightlike vectors such that $p_1\cdot p_2\not=0$, and ${\bf
  k}_i$ are transverse vectors, ${\bf k}_i\cdot p_j=0$ for all $i,j$.
The gluons have virtualities
\beq\label{virtcond}
k^2_i={\bf k}_i^2=-|{\bf k}_i|^2.
\eeq

The cross section $\sigma_{\rm off}(gg\to
H)$  is computed averaging over incoming
and summing over outgoing spin and color:
\beq\label{xsecrec}
\sigma_{\rm off} =\frac{1}{J}\frac{1}{256} 
{\cal M}_{ab}^{\mu\nu} {{\cal
    M}^*}_{ba}^{\mu^\prime\nu^\prime}\sum_{\lambda_1}\varepsilon^{\lambda_1}_\mu(k_1){\varepsilon^*}^{\lambda_1}_{\mu^\prime}(k_1)\sum_{\lambda_2}
\varepsilon^{\lambda_2}_\nu(k_2){\varepsilon^*}^{\lambda_2}_{\nu^\prime}(k_2)\,
 d{\cal P},
\eeq
where the flux factor 
\beq\label{flux}
J=2 (k_1\cdot k_2-{\bf  k}_1\cdot
    {\bf k}_2)
\eeq
is  determined on the surface orthogonal to
  $p_1,\>p_2$ eq.~(\ref{sudakov}), and
the phase space is
\beq\label{phase}
d{\cal P}= \frac{2\pi}{\mhiggs^2}
\delta\left(\frac{1}{z}-1-\frac{|{\bf k}_1+{\bf k}_2|^2}{m_H^2}\right).
\eeq
Note that the kinematics for a $2\to1$ process is fixed, so
eq.~(\ref{xsecrec}) gives the total cross section and  no phase--space
integration is needed. 

The sums over gluon polarizations are given by 
\beq\label{glupol}\sum_{\lambda_i}\varepsilon^{\lambda_i}_\mu(k_i)
{\varepsilon^*}^{\lambda_i}_\nu(k_i) =2 \frac{{\bf k}^\mu_i{\bf
    k}^\nu_i}{|{\bf k^2_i|}};\quad i=1,2.
\eeq
Here, the virtualities will be parametrized through  the dimensionless
variables
\beq\label{xidef}
\xi_i\equiv \frac{| {\bf k}_i|^2}{\mhiggs^2}.
\eeq
%Momentum conservation then implies 
%\beq\label{momcons}
%2 k_1\cdot k_2= \mhiggs^2\left(1+\xi_1+\xi_2\right).
%\eeq

The reduced cross section $\bar\sigma$, obtained extracting an overall
factor $\mhiggs^2$, 
\beq\label{redxdef}
\mhiggs^2 \sigma_{\rm off}(gg\to
H)\equiv\bar\sigma(\ytop;\xi_1,\xi_2,\varphi,z),
\eeq
is then a dimensionless function $\bar\sigma(\ytop;\xi_1,\xi_2,\varphi,z)$ 
of the parameter $\ytop$
eq.~(\ref{zetdef}) and of the kinematic variables
 $\xi_1$, $\xi_2$, the
relative  angle $\varphi$ of the two transverse momenta
\beq
\label{phidef}
\varphi=\cos^{-1}\left(\frac{{\bf k}_1\cdot {\bf k}_2}{|{\bf k}_1||{\bf k}_2|}\right),
\eeq
and
\beq\label{zdef}
z\equiv\frac{\mhiggs^2} {2 z_1z_2p_1\cdot p_2 }= 
\frac{\mhiggs^2}{2 \left(k_1\cdot k_2-{\bf k_1}\cdot {\bf
      k_2}\right)}.
\eeq
Note that, in the collinear limit ${\bf k}_1,\, {\bf k}_2\to0$,  $z$
eq.~(\ref{zdef}) reduces to $\tau$ eq.~(\ref{ydef}).

\item The reduced cross section is averaged  over $\varphi$, and its
  dependence on $z$ eq.~(\ref{zdef}) is Mellin-transformed:
\beq\label{melxsec}
\bar\sigma(N,\xi_1,\xi_2)=\int_0^1\!dz\,
z^{N-1}\int_0^{2\pi}\!\frac{d\varphi}{2\pi}\, 
\bar\sigma(\ytop;\xi_1,\xi_2,\varphi,z).
\eeq
\item The dependence on $\xi_i$ is also Mellin-transformed, and the
  coefficient of the collinear pole in $M_1$, $M_2$ is extracted:
\beq\label{mmel}
h(N,M_1,M_2)=M_1M_2 \int_0^\infty\! d\xi_1\,\int_0^\infty\! d\xi_2\,
\xi_1^{M_1-1} \xi_2^{M_2-1} \bar\sigma(N,\xi_1,\xi_2).
\eeq
Note that the integral in eq.~(\ref{mmel}) has a simple pole in
both $M_1=0$ and $M_2=0$. The  residue of this pole
is the usual hard coefficient function as
determined in collinear factorization, which is thus
$C(N)=h(N,0,0)$.
\item The leading singularities of the hard coefficient function
  eq.~(\ref{cmeldef}) are obtained by expanding in powers of $\as$ at
  fixed $\as/N$  the
  function obtained when  $M_1$ and $M_2$  in eq.~(\ref{mmel})
  are identified with the leading singularities of the largest eigenvalue of the
  singlet anomalous dimension matrix, namely
\beq\label{lsingcf}
\mhiggs^2\sigma_0(\ytop)
C(\as(\mhiggs^2),N,\ytop)
=h\left(N,\gamma_s\left(\frac{\as}{N}\right),\gamma_s\left(\frac{\as}{N}\right)
\right)\left[1+O\left(\as\right)\right].
\eeq
Here,  $\gamma_s$ is the leading order term in the expansion of 
the large eigenvalue $\gamma^+$ of the singlet anomalous
dimension matrix in powers of $\as$ at fixed $\as/N$:
\beq\label{lsingad}
\gamma^+(\as,N)=\gamma_s\left(\frac{\as}{N}\right)+\gamma_{ss}\left(\frac{\as}{N}\right)+\dots
,
\eeq
\end{itemize}
with~\cite{jaroszewicz}
\beq\label{gammas}
\gamma_s\left(\frac{\as}{N}\right)=\sum_{n=1}^\infty
c_n\left(\frac{C_A\as}{\pi N}\right)^n;\quad c_n=1,0,0,2\zeta(3),\dots
,
\eeq
where $C_A=3$.

So far, this procedure has been used  to determine the leading nontrivial
singularities to the hard coefficients for a small number of processes:
heavy quark photo-- and
electro--production~\cite{catani}, deep--inelastic
scattering~\cite{ch}, heavy quark hadroproduction~\cite{ellis,camici},
and Higgs production in the infinite $\mtop$ limit~\cite{hautmann}.

\subsection{Cross section for Higgs production from two off-shell gluons}
The leading--order amplitude for the production of a Higgs in the
fusion of two off--shell gluons with momenta $k_1$ and $k_2$ and color
$a$,$b$ is given
by the single triangle diagram, and it is equal to
\bea
  {\cal M}_{ab}^{\mu\nu} &=& 4i\,\delta^{ab}\,\frac{g_s^2\mtop^2} {v}\biggl[
      \frac{ k_2^{\mu} k_1^{\nu}}{\mhiggs^2} A_1(\xi_1,\xi_2;\ytop) -
        g^{\mu\nu} A_2(\xi_1,\xi_2;\ytop) \nonumber\\
  && + \left(\frac{k_1\cdot k_2}{\mhiggs^2} A_1(\xi_1,\xi_2;\ytop) -
    A_2(\xi_1,\xi_2;\ytop)\right) 
     \frac{ k_1\cdot k_2 k_1^{\mu} k_2^{\nu} - k_1^2 k_2^{\mu}
        k_2^{\nu} - k_2^2 k_1^{\mu} k_1^{\nu}}{k_1^2k_2^2} \biggr] \,,
\label{Dj}
\eea
where the strong coupling is $\alpha_s=\frac{g_s^2}{4\pi}$ and the
top Yukawa coupling is given by $h_t=\frac{m_t}{v}$ in terms of the Higgs
vacuum-expectation value $v$, related to the Fermi coupling by
$G_F=\frac{1}{\sqrt{2}v^2}$.
The dimensionless form factors $A_1(\xi_1,\xi_2;\ytop)$ and
$A_2(\xi_1,\xi_2;\ytop)$ have been computed in
ref.~\cite{delduca}; their explicit expression is given in the
appendix. They were subsequently rederived in Ref.~\cite{pasechnik},
where an expression for the Higgs production cross section from the
fusion of two off-shell gluons was also determined, but was not used to
obtain the high energy corrections to perturbative coefficient
functions.

The spin- and colour-averaged reduced cross section
eq.~(\ref{redxdef})
is then found
using eq.~(\ref{xsecrec}), with the phase space eq.~(\ref{phase}).
We get
\beq\label{osxs}
\bar\sigma(\ytop;\xi_1,\xi_2,\varphi,z) =8\sqrt{2}
 \pi^3\as^2 G_F\mhiggs^2\frac{\ytop^2}{\xi_1\xi_2} \left|\frac{1}{2 z}
   A_1 - A_2 \right|^2
\delta \left(\frac{1}{z}-1-\xi_1-\xi_2-\sqrt{\xi_1\xi_2}\cos\varphi\right).
\eeq

Because of the momentum--conserving delta, the Mellin transform  with
respect to $z$ is trivial, and the reduced cross section
eq.~(\ref{melxsec}) is given by
\bea \label{dimless}
&&\bar\sigma (N,\xi_1,\xi_2) =    8\sqrt{2}
 \pi^3\as^2 G_F\mhiggs^2\ytop^2
\int_{0}^{2 \pi} \frac{d \varphi}{2 \pi}
\frac{1}{(1+\xi_1+\xi_2)^N}\frac{1}{(1+ \sqrt{\alpha} \cos \varphi)^{N}} \\ \nonumber
&& \left[ \left( |{A}_1|^2 \cos^2 \varphi + \xi_1 \xi_2
  |{A}_3|^2 \right) 
+ \frac{1}{\sqrt{\xi_1 \xi_2}} \left[|{A}_1|^2 (1+ \xi_1+\xi_2) - ({A}_1^*{A}_2+{A}_1 {A}_2^*)\right] \cos \varphi \right] \,,
\eea
where we have defined the dimensionless variable
\beq
\alpha \equiv\frac{4 \xi_1 \xi_2}{(1+\xi_1+\xi_2)^2}.\label{aldef}
\eeq

The three form factors $A_i$ are independent of $\varphi$, so all the
angular integrals can be performed in  terms of hypergeometric
functions, with the result
\bea\label{angav}
&&
\bar\sigma (N,\xi_1,\xi_2)=  8\sqrt{2}
 \pi^3\as^2 G_F\mhiggs^2\ytop^2
\frac{1}{(1+\xi_1+\xi_2)^N}   \Bigg \{ \frac{|{A}_1|^2}{2} \Big({}_2F_1(\frac{N}{2},\frac{N+1}{2},2,\alpha)
 \nonumber \\
&&\quad
 +
\frac{\alpha}{4}N(N+1)
{}_2F_1(\frac{N+2}{2},\frac{N+3}{2},3,\alpha)\Big) +\xi_1 \xi_2
|{A}_3|^2  {}_2F_1(\frac{N}{2},\frac{N+1}{2},1,\alpha)
 \nonumber\\
&&\quad  
- N \left[|{A}_1|^2 (1+ \xi_1+\xi_2) - ({A}_1^*{A}_2+{A}_1 {A}_2^*)\right]     \frac{1}{1+\xi_1+\xi_2}
{}_2F_1(\frac{N+1}{2},\frac{N+2}{2},2,\alpha) \Bigg \} \,.
\eea
In the limit $\mtop\to\infty$, using the behaviour of the form factors
eq.~(\ref{a1a2t}) the term in square brackets in
eq.~(\ref{angav}) as well as the term proportional to $A_3$  are seen
to vanish. The remaining terms, proportional to $A_1$, give the result
in the pointlike limit. The reduced cross section in this limit was
already derived in ref.~\cite{hautmann} (see eq.~(9) of that
reference): our result differs from  that of ref.~\cite{hautmann},
though the disagreement is by terms of relative $O(N)$, hence it is
immaterial for the subsequent determination of the leading
singularities of the hard coefficient function.

\subsection{High energy behaviour}

The leading singularities of the coefficient function can now be
determined from the  Mellin transform $h(N,M_1,M_2)$
eq.~(\ref{mmel}) of
the reduced cross section eq.~(\ref{angav}),  letting
$M_1=M_2=\gamma_s(\as/N)$ according to 
 eq.~(\ref{lsingcf}), and expanding in powers of $\as$ (i.e. effectively
 in powers of $M_1$, $M_2$) and then in
powers of $N$ about $N=0$. In the pointlike case ($\mtop\to\infty$)
the
Mellin integral eq.~(\ref{mmel}) diverges for all $M_1,\,M_2$ when
$N=0$, and it only has a region of convergence when $N>0$. As a
consequence, the function $h(N,M_1,M_2)$ eq.~(\ref{mmel}) has
singularities in the $M$ plane whose location depends on the value of
$N$, namely, simple poles of the form $\frac{1}{N-M_1-M_2}$: the
expansion in powers of $M_i$ has finite radius of convergence $M_i<N$,
leading to an expansion in powers of $\frac{M_i}{N}$ and thus double
poles when $M_i=\gamma_s$. 

In the resolved case (finite $\mtop$) we expect the Mellin integral to
converge when $N=0$ at least for $0< M_i<M_0$, for some real positive
$M_0$. We can then set $N=0$, and obtain the leading singularities of
the coefficient function from the expansion in powers of $M$ of
$h(0,M,M)$, letting $M=\gamma_s$. This turns out to be indeed the
case: when $N=0$,  $\bar\sigma(N,\xi_1,\xi_2)$
 only depends on $\xi_1,\,\xi_2$ through the form factors, and 
the combination of form
factors  which appear in $\bar\sigma$ eq.~(\ref{angav}) is regular
when $\xi_1,\,\xi_2\to0$ (see
 eq.~(\ref{osas})), while it vanishes when  $\xi_1,\,\xi_2\to\infty$
 (see eq.~(\ref{hetriv})).  Hence, we can let $N=0$ in  $\bar
 \sigma$, and get
\bea \label{impact}
&&h(0,M_1,M_2)=    8\sqrt{2}
 \pi^3\as^2 G_F\mhiggs^2\ytop^2\nonumber\\
&&\qquad\quad
\times M_1 M_2 \int_0^{+\infty}d \xi_1 \xi_1^{M_1-1}\int_0^{+\infty}d \xi_2
\xi_2^{M_2-1} 
\left[\frac{1}{2} |{A}_1|^2 + \xi_1 \xi_2 |{A}_3|^2\right].
\eea

Because the term in square brackets in eq.~(\ref{impact})  tends to a
constant as $\xi_1,\,\xi_2\to0$, the integrals in
eq.~(\ref{impact}) have an isolated  simple pole in $M_1$ and $M_2$, 
and thus the
Taylor expansion of $h(N,M_1,M_2)$ has a finite radius of
convergence. We can then determine the Taylor coefficients by expanding the
integrand of eq.~(\ref{impact}) and integrating term by term. It
follows from eqs.~(\ref{lsingcf}-\ref{gammas}) that knowledge of the
coefficients up to $k$-th order in both $M_1$ and $M_2$ is necessary
and sufficient to determine the leading singularity of the coefficient
function up to order $\as^k$.

Let us now determine the leading singularities of first three coefficients of the
expansion of the coefficient function eq.~(\ref{cfdef}). 
The constant term determines the leading--order result
$\sigma_0$ eq.~(\ref{cfdef}):
\beq\label{sigzerodet}
\mhiggs^2\sigma_0(\ytop)=h(0,0,0).
\eeq
Using the on-shell limit of the form factors (see eq.~(\ref{osas}) of
the appendix) in eq.~(\ref{impact}) we  reproduce the well-known result
eq.~(\ref{sigzero}).

The next-to-leading order term $C^{(1)}(N,\ytop)$ is determined by noting that
\begin{eqnarray}\label{hfirstord}
 h(0,M,0)&=&  4\sqrt{2}
 \pi^3\as^2 G_F\mhiggs^2\ytop^2 M\int_0^{+\infty} d \xi\, \xi^{M-1} |{A}_1(\xi,0)|^2\nonumber\\
 &=& h(0,0,0)-   8\sqrt{2}
 \pi^3\as^2 G_F\mhiggs^2\ytop^2
M\int_0^{+\infty} d \xi \ln \xi \frac{d |{A}_1(\xi,0)|^2}{d \xi}+ O(M^2)\,.
\end{eqnarray}
Equations~(\ref{lsingcf}-\ref{gammas}) then immediately imply that
\bea
\label{nlores}
&&C^{(1)}(N,\ytop)={\mathcal
    C}^{(1)}(\ytop) \frac{C_A}{N}\left[1+O(N)\right],\nonumber\\
&&\quad{\mathcal
    C}^{(1)}=-\frac{2 (8 \pi^2)^2 }{\left|\left(1- \frac{1}{4} (1-4 \ytop)
 s_0(\ytop)^2\right)\right|^2} \int_0^{+\infty} d \xi \ln \xi \frac{d |{A}_1(\xi,0)|^2}{d \xi} .
\eea

The value of the coefficient ${\mathcal
    C}^{(1)}$, determined from a numerical evaluation of the integral
in eq.~(\ref{nlores}), is tabulated in table~1  as a function of the
Higgs mass. Upon inverse Mellin
transformation, one finds that
\beq\label{nlosx}
\lim_{\tau\to 0} C^{(1)}(\tau,\ytop)= C_A {\mathcal
    C}^{(1)}(\ytop).
\eeq
The values given in  table~1  are indeed found to  
be in perfect agreement with a numerical evaluation of the  small
$\tau$ 
limit of the 
full NLO coefficient function $C^{(1)}(\tau,\ytop)$~\cite{spira}, for which we
have used  the form given in
ref.~\cite{bonciani}.

  \begin{table}[t!] \label{tabNNLO}
  \begin{center}
 \begin{tabular}{|c|c|c|}
 \hline
  $m_H$ & ${\mathcal C}^{(1)}(\ytop)$ & ${\mathcal C}^{(2)}(\ytop)$ \\ 
 \hline
  110 &  5.0447 &  16.2570  \\
  120 &  4.6873 &  14.5133  \\
  130 &  4.3568 &  13.0155  \\
  140 &  4.0490 &  11.7196   \\
  150 &  3.7607 &  10.5919   \\
  160 &  3.4890 &   9.6058  \\
  170 &  3.2318 &   8.7406 \\
  180 &  2.9872 &   7.9794 \\
  190 &  2.7536 &   7.3085 \\
  200 &  2.5296  &  6.7166 \\
  210 &  2.3140 &   6.1946 \\
  220 &  2.1057  &  5.7346 \\
  230 &  1.9037  &  5.3303 \\
  240 &  1.7072 &   4.9761 \\
  250 &  1.5151 &   4.6677 \\
  260 &  1.3267 &   4.4013 \\
  270 &  1.1409 &   4.1738 \\
  280 &  0.9568 &   3.9828 \\
  290 &  0.7731 &   3.8268 \\
  300 &  0.5884 &   3.7049 \\
 \hline
  \end{tabular}
   \caption{Values of the coefficients eq.~(\ref{nlores}) and
     eq.~(\ref{nnlores}) of the $O(\as/N)$ and  $O((\as/N)^2)$ of the  
     leading singularities of the coefficient function
     $C(\as(\mhiggs^2);N,\ytop)$ eq.~(\ref{cmeldef}).}
   \end{center}
  \end{table}
Turning finally to the determination of the hitherto unknown NNLO
leading singularity, we evaluate the $O(M^2)$ 
terms in the expansion eq.~(\ref{hfirstord}): by using again  
eqs.~(\ref{lsingcf}-\ref{gammas}) we find
\bea\label{nnlores}
&&C^{(2)}(N,\ytop)= {\mathcal C}^{(2)}(\ytop) \frac{
 C_A^2  }{N^2}\left[1+O(N)\right]\nonumber\\
&&\quad{\mathcal
    C}^{(2)}(\ytop)=
-\frac{(8\pi^2)^2}{\left|\left(1- \frac{1}{4}
(1-4 \ytop)
 s_0(\ytop)^2\right)\right|^2}\left\{\int_0^{+\infty} d \xi
\ln^2 \xi \frac{d |{A}_1(\xi,0)|^2}{d \xi}\right.\\
&&\qquad\qquad\left.-\int_0^{+\infty}d \xi_1 \int_0^{+\infty}d \xi_2 \left[ \ln \xi_1 \ln \xi_2
 \frac{\partial^2 |{A}_1(\xi_1,\xi_2)|^2}{\partial \xi_1\partial \xi_2} + 2 |{A}_3(\xi_1,\xi_2)|^2\right]\right\}.\nonumber
 \eea

The value  of the NNLO coefficient ${\mathcal
    C}^{(2)}(\ytop)$ obtained from numerical
evaluation of the integrals in eq.~(\ref{nnlores}) is also tabulated  
in  table~1. This is the main result of the present paper.

\section{Improvement of the NLO and NNLO cross sections}

Knowledge of the leading small $\tau$ behaviour of the exact coefficient
function $C(\as(\mhiggs^2);\tau,\ytop)$ eq.~(\ref{cfdef}) can be
used to improve its determination. 
Indeed, as discussed in section~1, we expect the pointlike
($\mtop\to\infty$) approximation to be quite accurate at large $\tau$,
whereas we know that it must break down as $\tau\to0$. Specifically, 
 the small $\tau$ behaviour of the coefficient function is
dominated by the highest rightmost singularity in  $C(\as(\mhiggs^2);N,\ytop)$
eq.~(\ref{cmeldef}), which for the exact result is  a $k$-th order pole but
becomes a   $2k$-th order pole in the pointlike approximation. Hence
the pointlike approximation displays a spurious stronger growth
eq.~(\ref{hebeh}) at small enough $\tau$. 

Having determined the exact small
$\tau$ behaviour up to NNLO, we can improve the  approximate pointlike determination of the
coefficient function by subtracting its  spurious small $\tau$ growth and
replacing it with the exact behaviour.
We discuss first the NLO case, where the
full exact result is known, and then turn to the NNLO where only the
$\mtop\to\infty$ result is available.

\subsection{NLO results}
\begin{figure}[t]
\begin{center}
\epsfig{width=.7\textwidth, file=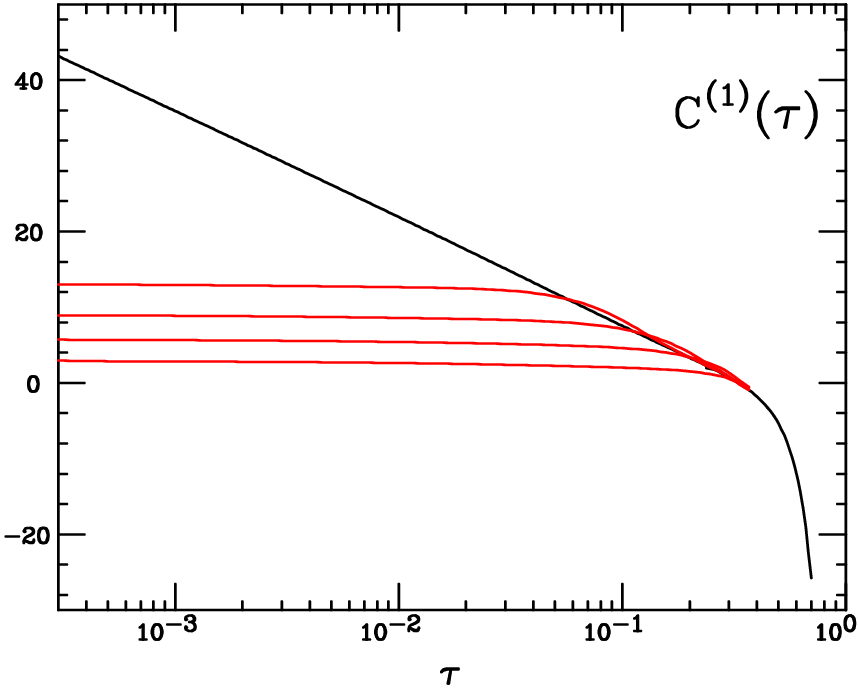}  
\end{center}
\begin{center} 
 \caption{The hard coefficient $C^{(1)}(\tau,\ytop)$ eq.~(\ref{cfexp})
 (parton--level coefficient function normalized to the Born result)
   plotted
     as a function of $\tau$. The curves from top to bottom on the left
     correspond to $\mtop=\infty$ (black), and to $\mtop=170.9$~GeV (red), with
     $\mhiggs=130,\>180,\>230,\>280$~GeV.
 \label{nlomt}}\end{center}
\end{figure}
At NLO the small $\tau$ behaviour of the coefficient function in the pointlike
approximation is
dominated by a double  pole, whereas it is given by  the simple pole
eq.~(\ref{nlores}) in the exact case. This
corresponds to an exact NLO contribution $C^{(1)}(\tau,\ytop)$
which tends to a constant at small $\tau$, whereas the pointlike
approximation to it grows as $\ln \tau$:
\bea\label{cnlopoint}
C^{(1)}(\tau,\infty)&=& d^{(1)}_{\rm
  point}(\tau)+O\left(\tau\right);\quad d^{(1)}_{\rm point}(\tau)=  c^1{}_2 \ln \tau+ c^1{}_1 \\
\label{cnloex}
C^{(1)}(\tau,\ytop)&=& d^{(1)}_{\rm
  ex}(\tau,\ytop)+O\left(\tau\right);\quad d^{(1)}_{\rm ex}(\tau,\ytop) =3{\mathcal
    C}^{(1)}(\ytop),
\eea
where ${\mathcal
    C}^{(1)}(\ytop)$ is tabulated in table~1, while from
  Refs.~\cite{spira,djouadi,dawson}
we get 
\beq\label{appco}
c^1{}_2 =-6;\quad c^1{}_1=-\frac{11}{2} .
\eeq

The  NLO term
$C^{(1)}(\tau,\ytop)$
eq.~(\ref{cfdef}) is plotted  as a function of $\tau$ in
fig.~\ref{nlomt}, both in the pointlike ($\mtop\to\infty$)
approximation, and in its exact form computed with increasing values
of the Higgs mass, i.e. decreasing values of $\ytop$. It is apparent
that the pointlike approximation is very accurate, up to the point
where the spurious logarithmic growth eq.~(\ref{cnlopoint}) sets in.

\begin{figure}[t]
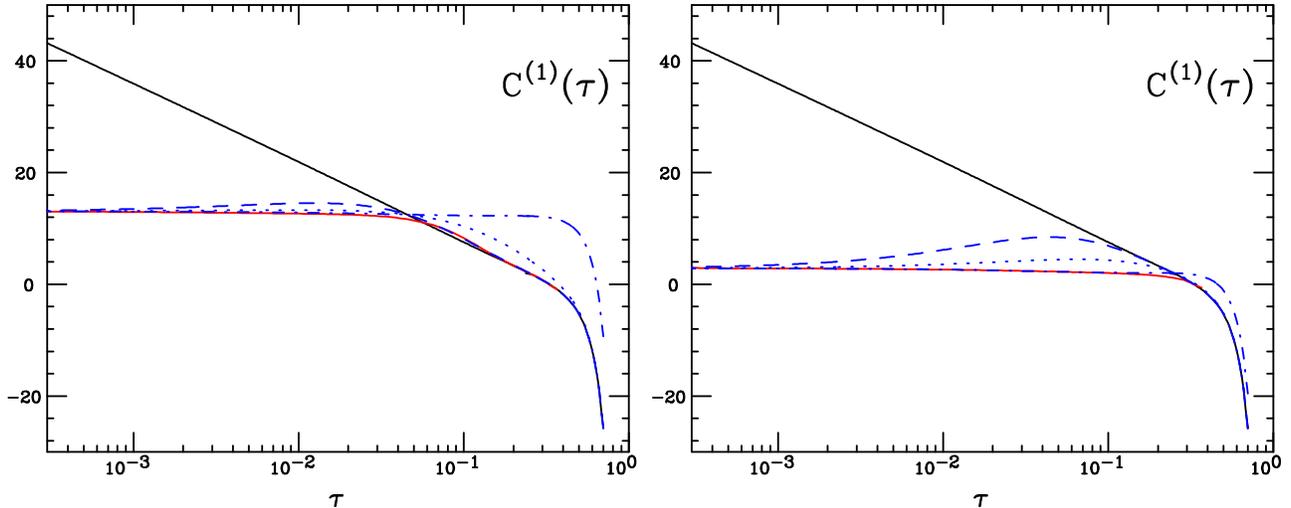

\begin{center}
\epsfig{width=.49\textwidth, file=nlo130.ps}  
\epsfig{width=.49\textwidth, file=nlo280.ps}  
\end{center}
\begin{center} 
 \caption{The hard coefficient $C^{(1)}(\tau,\ytop)$
     eq.~(\ref{cfexp}) with $\mhiggs=130$~GeV (left) and
     $\mhiggs=280$~GeV (right). 
  The solid curves
     correspond to $\mtop=\infty$ (black) and $\mtop=170.9$~GeV (red),
     (same as fig.~\ref{nlomt}). The three blue curves correspond to the
     approximation eq.~(\ref{nloappr},\ref{tchoice}), with $k=0$
     (dot-dashed), $k=5$ (dotted), $k=20$ (dashed).
 \label{nlo130280}}\end{center}
\end{figure}
We can construct an approximation to $C^{(1)}(\tau,\ytop)$ by
combining the pointlike approximation with the exact small $\tau$
behaviour:
\beq
\label{nloappr}
C^{(1), {\rm app.}}(\tau,\ytop)\approx C^{(1)}(\tau,\infty)+\left[ d^{(1)}_{\rm
    ex}(\tau,\ytop) -d^{(1)}_{\rm point}(\tau)
\right]T(\tau)
\eeq
where $d^{(1)}_{\rm
    ex}(\tau,\ytop)$ and $d^{(1)}_{\rm point}(\tau)$ are defined as in
  eq.~(\ref{cnloex}) and eq.~(\ref{cnlopoint}) respectively, while
  $T(\tau)$ is an interpolating function, which we may introduce in
  order to tune the point where the small $\tau$ behaviour given by 
$d^{(1)}_{\rm
    ex}(\tau,\ytop)$ sets in. Clearly, as $\tau\to0$ the approximation
  eq.~(\ref{nloappr}) reproduces the exact small $\tau$ behaviour of the
  exact coefficient function eq.~(\ref{cnloex}), provided only the
  interpolating function $\lim_{\tau\to  0} T(\tau)=1$. Furthermore,  as discussed in section~1, the
  behaviour of the coefficient function $C^{(1)}(\tau,\ytop)$  as 
 $\tau\to 1$  is to all orders controlled by soft gluon radiation, which leads to
 contributions to $C^{(1)}(\tau,\ytop)$  which do not depend on
 $\ytop$ and diverge as $\tau\to 1$. Hence, the pointlike approximation
 is exact as $\tau\to1$. 
Because the functions  $d^{(1)}_{\rm
    ex}(\tau,\ytop)$ and $d^{(1)}_{\rm point}(\tau)$ are regular as $\tau\to1$,
  this exact behaviour is also reproduced by the approximation
  eq.~(\ref{nloappr}), provided only  $\lim_{\tau\to  1} T(\tau)$
  is finite. Hence, $C^{(1), {\rm app.}}(\tau,\ytop)$ reproduces the
  exact $C^{(1)}(\tau,\ytop)$ as $\tau\to0$ up to terms that vanish as
  $\tau\to0$ and as $\tau\to 1$ up to terms that are nonsingular as
  $\tau\to1$, even when $T(\tau)=1$. 

Nevertheless, we may also choose $T(\tau)$ in
  such a way that $T(1)=0$ (while $T(0)=1$ always), so that
  $C^{(1)}(\tau,\ytop)$  agrees with the pointlike approximation
  $C^{(1)}(\tau,\infty)$ in some neighborhood of $\tau=1$. 
%For example,
%  fig.~1 suggests that we
%  let $T(\tau)=\Theta(\tau_0-\tau)$, where $\Theta(\tau)$ is the Heaviside
%  function and $0<\tau_0<1$, so that $C^{(1), {\rm app.}}(\tau,\ytop)$
%  only differs from the pointlike approximation when $\tau<\tau_0$, and then
%  for each $\ytop$ we choose $\tau_0$ as the value of $\tau$  such that the pointlike
%  approximation equals the exact asymptotic small $\tau$ value:
%  $C^{(1)}(\tau,\tau_0)=d^{(1)}_{\rm ex}(0,\ytop)$. This choice however
%  leads to a  form of  $C^{(1), {\rm app.}}(\tau,\ytop)$ whose first
%  derivative is discontinuous at $\tau=\tau_0$. 
For instance, we can let 
\beq\label{tchoice}
T(\tau)=(1-\tau)^k,
\eeq
with $k$ real and positive, so that the first $k$ orders of the Taylor
expansion about $\tau=1$ of $C^{(1), {\rm app.}}(\tau,\ytop)$ and the
pointlike approximation coincide. By varying the value of $k$, we can
choose the matching point $\tau_0$, such that
$C^{(1), {\rm app.}}(\tau,\ytop)$ only differs significantly from
the pointlike approximation if $\tau<\tau_0$: a larger value
of $k$ leads to a smaller value of $\tau_0$. 

In fig.~2 we compare
the approximate NLO term eq.~(\ref{nloappr}) to the exact
and pointlike results, for two different values of $\ytop$, with
$T(\tau)$ given by eq.~(\ref{tchoice})
and a choice of $k$ which leads to different values of the matching
between approximate and pointlike curves. It appears that an optimal
matching is obtained by choosing $k$  in such a way that the
approximation eq.~(\ref{nloappr}) matches the pointlike result close
to the point
where the logarithmic growth of the latter intersects  the asymptotic
constant value of the exact result. Note that this optimal matching
could be determined without knowledge of the exact result.
With this choice, the
approximation eq.~(\ref{nloappr}) differs from the exact result for
the NLO contribution to the partonic cross section by less than $5\%$
for all values of $\tau$.

\subsection{NNLO and beyond}
\begin{figure}[t]
\begin{center}
\epsfig{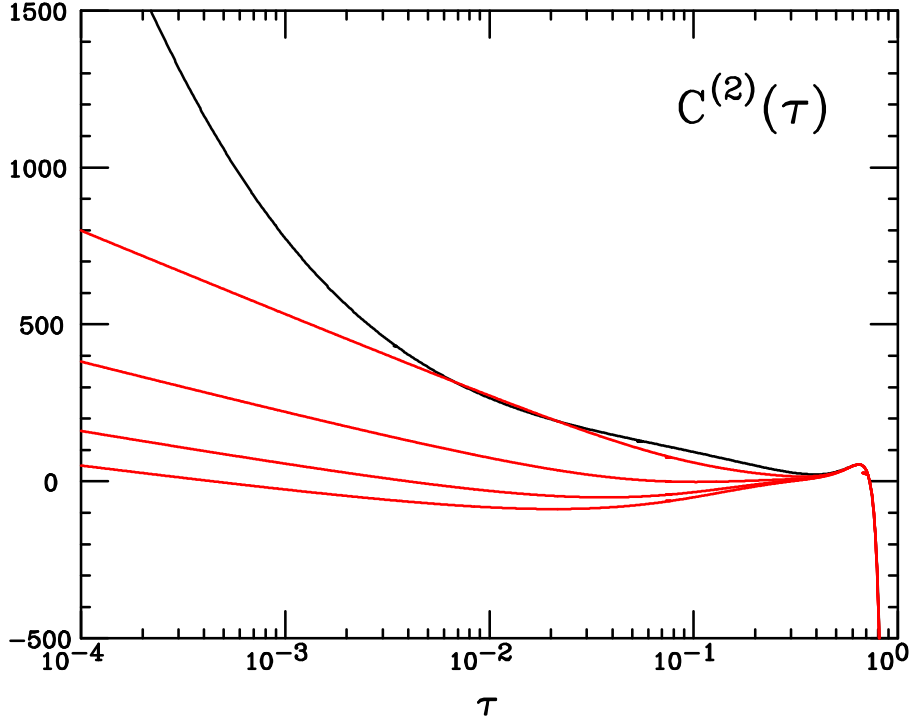}  
\end{center}
\begin{center} 
 \caption{The hard coefficient $C^{(2)}(\tau,\ytop)$ eq.~(\ref{cfexp})
   (parton--level coefficient function normalized to the Born result)
   plotted  as a function of $\tau$. The curves from top to bottom on the left
     correspond to $\mtop=\infty$ (black), and to the approximation
 eq.~(\ref{nnloappr}) with $T(\tau)$ eq.~(\ref{tchoice}) and $k=5$, and
 $\mtop=170.9$~GeV (red), with
     $\mhiggs=130,\>180,\>230,\>280$~GeV.
 \label{nnlomt}}\end{center}
\end{figure}
At NNLO, the pointlike approximation to the coefficient function has a
quadruple pole at $N=0$, corresponding to a  $\ln^3 \tau$ rise, while the exact
result only has a double pole, and thus it rises linearly with $\ln
\tau$:
\bea\label{cnnlopoint}
C^{(2)}(\tau,\infty)&=& d^{(2)}_{\rm
  point}(\tau)+O\left(\tau^0\right);\quad d^{(2)}_{\rm point}(\tau)=  c^2{}_4 \ln^3 \tau+ c^2{}_3 \ln^2 \tau+ c^2{}_2 \ln \tau \\
\label{cnnloex}
C^{(2)}(\tau,\ytop)&=& d^{(2)}_{\rm
  ex}(\tau,\ytop)+O\left(\tau^0\right);\quad d^{(2)}_{\rm ex}(\tau,\ytop) =-9\, {\mathcal
    C}^{(2)}(\ytop) \ln \tau ,
\eea
where ${\mathcal
    C}^{(2)}(\ytop)$ is tabulated in table~1, while from
  Ref.~\cite{anastasiou}
we get 
\beq\label{nnappco}
c^2{}_4 =-6;
\quad c^2{}_3 =-\frac{231}{4}+n_f\frac{17}{18} ;\quad c^2{}_2 =
\left(-\frac{2333}{8}+3 \pi^2\right)+n_f\frac{641}{108},
\eeq
where  $n_f$ the  number of
flavors. 

\begin{figure}[t]
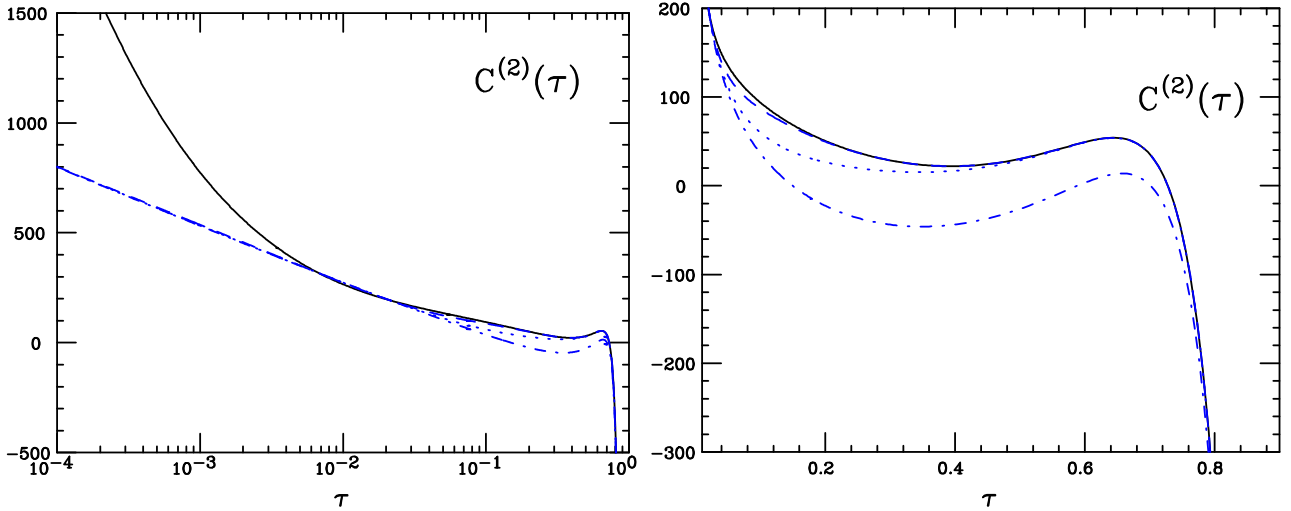

\begin{center}
\epsfig{width=.49\textwidth, file=nnlo130.ps}  
\epsfig{width=.49\textwidth, file=nnlo130det.ps}  
\end{center}
\begin{center} 
 \caption{The hard coefficient $C^{(2)}(\tau,\ytop)$
     eq.~(\ref{cfexp}) with $\mhiggs=130$~GeV, plotted versus $\tau$ on a
     logarithmic (left) or linear (right) scale.
  The solid black curve
     corresponds to $\mtop=\infty$ (black, same as fig.~\ref{nnlomt})), and the three blue curves 
 are the approximation eq.~(\ref{nnloappr}) with $\mtop=170.9$~GeV and $T(\tau)$ eq.~(\ref{tchoice}) with
 and $k=0$ (dot-dashed), $k=5$ (dotted, same as fig.~\ref{nnlomt}), $k=20$ (dashed).
 \label{nnlo130}}\end{center}
\end{figure}
At this order, the exact form of $C^{(2)}(\tau,\ytop)$ is not
known. However, analogously to the NLO case, we construct an approximation to it based on its
determination~\cite{anastasiou} in 
the pointlike limit, combined with the exact small $\tau$
behaviour eq.~(\ref{nnlores}):
\beq
\label{nnloappr}
C^{(2), {\rm app.}}(\tau,\ytop)\approx C^{(2)}(\tau,\infty)+\left[ d^{(2)}_{\rm
    ex}(\tau,\ytop) -d^{(2)}_{\rm point}(\tau)
\right]T(\tau)
\eeq
with  $d^{(2)}_{\rm
    ex}(\tau,\ytop)$ and $d^{(2)}_{\rm point}(\tau)$ defined  in
  eq.~(\ref{cnnloex}) and eq.~(\ref{cnnlopoint}) respectively, and 
  $T(\tau)$ an interpolating function as discussed in section~3.1.
Note that as $\tau\to0$ the approximation eq.~(\ref{nnloappr}) only
reproduces the exact result up to a constant, whereas at NLO the
approximation eq.~(\ref{nloappr}) reproduces the exact result up to
terms which vanish at least as $O(\tau)$ .

\begin{figure}[t]
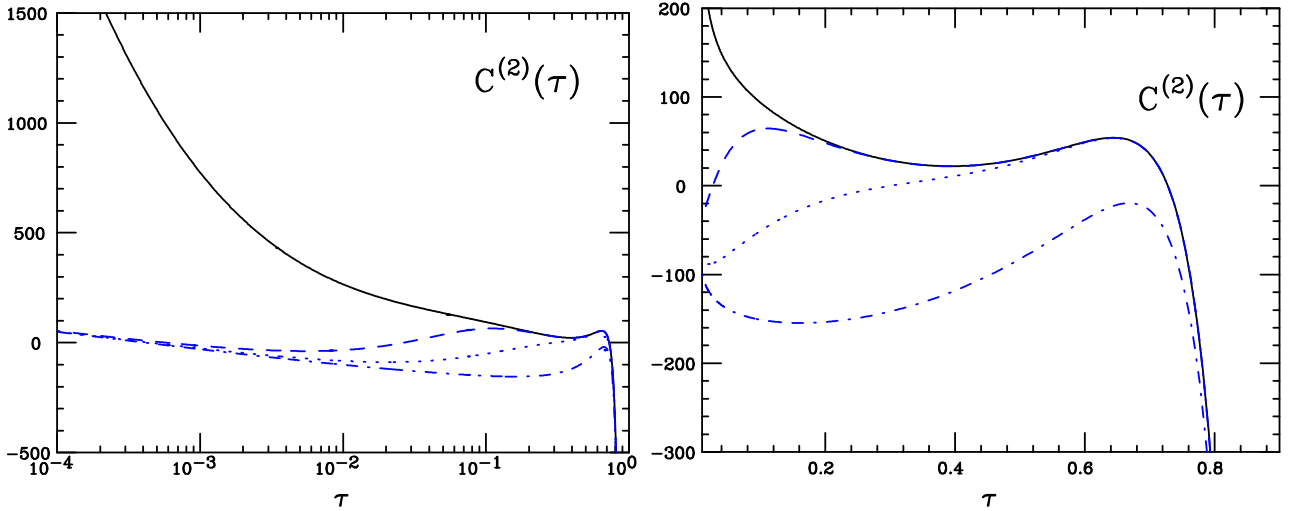

\begin{center}
\epsfig{width=.49\textwidth, file=nnlo280.ps}  
\epsfig{width=.49\textwidth, file=nnlo280det.ps}  
\end{center}
\begin{center} 
 \caption{Same as fig.~\ref{nnlo130}, but now with
$\mhiggs=280$~GeV.
 \label{nnlo280}}\end{center}
\end{figure}
The approximation to the exact result $C^{(2), {\rm
    app.}}(\tau,\ytop)$, computed using
${\mathcal C}^{(2)}$ from table~1 with four different values of the
Higgs mass, and taking $T(\tau)$ eq.~(\ref{tchoice})
with $k=5$ 
is compared in fig.~3 to the pointlike approximation $C^{(2), {\rm
    app.}}(\tau,\ytop)$ of ref.~\cite{anastasiou} (with
$n_f=5$). 
In figures~4-5 we further compare the results obtained with different
choices of the matching function $T(\tau)$ eq.~(\ref{tchoice}), and the same
two values of the Higgs mass used to produce figs.~2-3 at NLO. 

At this order, the contribution from the leading small $\tau$
logs to the pointlike $C^{(2), {\rm
    app.}}(\tau,\infty)$ is sizable even for large $\tau$. Indeed,
figs.~4-5 show that 
the behaviour of $C^{(2)}$  around its local maximum 
at $\tau\approx0.65$ receives a sizable contribution from 
 the $\ln\tau$ rise and
$\ln^2\tau$ drop eq.~(\ref{cnnlopoint}).  If these are removed by using 
eqs.~(\ref{nnloappr},\ref{tchoice}) with $k=0$, the shape of  $C^{(2)}$
around the maximum is affected significantly, but if the matching is moved to
smaller $\tau$ by choosing $k\gsim 5$ the maximum is reproduced.
Hence, whereas we can still obtain a rather smooth matching  at any
desired value of $\tau$ the choice of the optimal value of $\tau$ is
not obvious. In particular, matching at a  value of $\tau$ where
the contribution of the asymptotically spurious $\ln^2\tau$ becomes 
significant  leads to rather large values of the matching point
$\tau\gsim0.6$. 
Anyway, it is clear that the pointlike approximation
breaks down for $\tau\lsim 0.1$.

Contributions beyond NNLO in the expansion of
$h(N,\gamma_s,\gamma_s)$ eq.~(\ref{lsingcf}) in powers of
$\frac{\as}{N}$ can be determined by pursuing the expansion of
$h(0,M,M)$ eq.~(\ref{impact}) in powers of $M$, and determining
numerically the ensuing integrals, which have the form of
eqs.~(\ref{nlores},\ref{nnlores}), but with higher order powers of
$\ln\xi_1,\,\ln\xi_2$. The series of contributions to the
coefficient function eq.~(\ref{cfdef}) thus obtained has a finite radius of
convergence in $N$--Mellin space, dictated by the location of the
rightmost singularity in $\gamma_s$, and thus in $\tau$ space
it converges for all
$0<\tau\le 1$~\cite{summing}. Therefore, its resummation can be accomplished to
arbitrary accuracy by computation of a finite number of terms. This
resummation, however, induces spurious singularities in the $N$--space
coefficient function, which can be removed by the inclusion of a
suitable class of formally subleading running-coupling corrections, as
recently shown in Ref.~\cite{ball}.

\subsection{$K$ factors}
  \begin{table}[t!] \label{tabkfact}
  \begin{center}
 \begin{tabular}{|c|c|c|}
 \hline
   & $\kappa^{\rm NLO}$ & $\kappa^{\rm NNLO}$ \\ 
 \hline
\multicolumn{3}{|c|}{$\mhiggs=130$~GeV}\\
\hline
 pointlike   &  36.69 &  658   \\
 exact   &  36.58 &  n.a.  \\
appr., $k=5$  &  37.64 &  648   \\
appr., $k=20$   &  36.66 & 655     \\
 \hline
\multicolumn{3}{|c|}{$\mhiggs=280$~GeV}\\
\hline
 pointlike   &  38.08 &  716  \\
 exact   &  37.47 &  n.a.   \\
appr., $k=2$   & 37.97 &  670   \\
appr., $k=5$  &  37.73 &  693  \\
 \hline
  \end{tabular}
   \caption{The NLO and NNLO contributions to the $K$ factor 
 eq.~(\ref{kfactdef}), computed with
     center-of-mass energy $s=14$~TeV, and  $\mtop\to\infty$, denoted with
     pointlike, or $\mtop=170.9$~GeV, denoted with  exact or
     approximate. The approximate result uses
     eqs.~(\ref{nloappr},\ref{nnloappr}), with $T(\tau)$
     eq.~(\ref{tchoice}) and the value of $k$ given in the table. The
     MRST2002~\cite{mrst2002} gluon distribution has been used.
}
   \end{center}
  \end{table}
The accuracy of the various approximations at the level of hadronic
observables  clearly depends on the individual process. For the
total inclusive cross section eq.~(\ref{partohadr}), as is well known,
the pointlike approximation is actually very good, and thus the impact
of the improvement eq.~(\ref{nloappr}) is moderate. To give a
quantitative assessment, we define a $K$ factor by letting:
%NLO and NNLO $\kappa$ factors by letting
\bea\label{kfactdef}
\sigma_{gg}(\tau_h;\ytop,\mhiggs^2)&=&\sigma^{0}_{gg}(\tau_h;\ytop,\mhiggs^2) K(\tau_h;\ytop,\mhiggs^2)
\nonumber\\ 
K(\tau_h;\ytop,\mhiggs^2)&=&1+
\frac{\as(\mhiggs^2)}{\pi} \kappa^{\rm NLO}(\tau_h;\ytop,\mhiggs^2)+
\left(\frac{\as(\mhiggs^2)}{\pi}\right)^2 \kappa^{\rm
  NNLO}(\tau_h;\ytop,\mhiggs^2) \nonumber \\ && +\, O\left(\as^3(\mhiggs^2)\right),
\eea
where $\sigma^{0}_{gg}$ is the leading--order form of the contribution
eq.~(\ref{partohadr}) of 
the gluon--gluon channel to total
hadronic cross section. The value of the NLO and NNLO contributions
to the $K$ factors, determined
using the MRST2002~\cite{mrst2002} gluon distribution
in eq.~~(\ref{partohadr}) are given
in table~2 at LHC energies for two values of the Higgs mass, both in
the pointlike, exact and approximate (eq.~(\ref{nloappr}) and eq.~(\ref{nnloappr})) cases. 

At NLO with $\mhiggs=130$~GeV (``light''), the pointlike approximation
to $\kappa^{\rm NLO}$ deviates by
$0.3\%$ from the exact result, and even with  $\mhiggs=280$~GeV
(``heavy'') it
only deviates by $1.6\%$. It should be kept in mind, however, that 
$\kappa^{\rm NLO}$ itself is quite large: for $\as\approx 0.1$, it amounts to
a $\sim100$\% contribution to the $K$  factor 
eq.~(\ref{kfactdef}). Hence, the error made using the pointlike
approximation is between the per mille and the per cent level, and thus
not entirely negligible in a precision analysis.

 Using the approximation
eqs.~(\ref{nloappr}-\ref{tchoice}) with the values $k=20$ for light
Higgs and $k=5$ for heavy Higgs, which are seen from fig.~2 to give
good matching, the deviation can be reduced to $0.2\%$ and $0.7\%$
respectively, and even more accurate results could be obtained by an
optimization of the matching. However, a poor choice of the matching
(such as $k=5$ for light Higgs or $k=2$ for heavy Higgs) can lead to a
result at the hadronic level which is actually closer  to the
pointlike approximation, or even worse than it.
It is clear that at
the partonic level the small $\tau$ behaviour eq.~(\ref{cnlopoint})
accounts for most of the discrepancy between the exact and pointlike
results, and even the determination of
a hadronic observable which depends very little on
the parton-level small $\tau$ behaviour can be improved very
substantially for values of $\tau_H$ relevant for LHC by using the
approximation eq.~(\ref{nloappr}).

%Whereas a detailed study of the 
%phenomenological implications of this fact will be performed 
%elsewhere~\cite{marzani}, 
The NNLO contribution  $\kappa^{\rm NNLO}$ is not known. Its values
 computed in the pointlike
approximation, or with the approximation
eqs.~(\ref{nnloappr},\ref{tchoice}) and different choices of $k$ are
shown in table~2. Even
at the inclusive hadronic level, now the size of the NNLO contribution
can change up to about $5-10\%$ if the matching is performed at large
$\tau$. Furthermore,
$\kappa^{\rm NNLO}$  is also  quite large: with $\as\approx0.1$, 
it amounts to a $\sim50\%$ correction to the leading order, and thus
to a further $\sim25\%$ correction to the $K$ factor. Therefore,
 the impact of the pointlike approximation at NNLO is up to
several per cent of the total $K$ factor, rather larger that the
impact of the pointlike approximation at NLO, and comparable to
uncertainties which are currently discussed in precision studies at NNLO.

\section{Outlook}
In this paper we have determined the leading high energy
(i.e. small $\tau=\frac{\mhiggs^2}{\hat s}$) singularities 
 of the cross section for Higgs production in
gluon--gluon fusion to all orders in the strong coupling, by providing
an expression (eq.~(\ref{impact})) whence the coefficients of these
singularities can be obtained by Taylor expanding and computing a
double integral. We have given explicit numerical expressions for these
coefficients up to NNLO. 

The high energy behaviour of this
cross section is different according to whether it is determined with
finite $\mtop$ or with $\mtop\to\infty$ (pointlike
approximation). It turns out that at NLO this different
high energy behaviour is responsible for most of the
discrepancy between the pointlike approximation and the exact
result. As a consequence, an accurate approximation to the exact
result can be constructed by combining the pointlike approximation at
large $\tau$ with the exact small $\tau$ behaviour. Some care
must be taken in matching, but very accurate results can be obtained
by simply choosing the matching point as that where the spurious small
$\tau$ behaviour of the pointlike behaviour sets in.

At NNLO, where the exact result is not
known, the impact of the high energy behaviour turns out to be large even for
moderate values of $\tau\sim 0.5$. Hence, an approximation
constructed analogously to that which is successful at NLO, namely 
matching the pointlike limit to the asymptotic exact behaviour  at the
point where the asymptotically spurious terms become
significant, leads to an approximate result which 
differs significantly from the pointlike approximation for most values of the
partonic center-of-mass energy.

The effect of these high energy terms on the total inclusive hadronic
cross section remains quite small, because the latter is dominated by
the region of low partonic center--of--mass energy, partly due to 
shape of the gluon parton distributions, which are peaked in the
region where the gluons carry a small fraction of the incoming
nucleon's energy, and partly because the partonic cross section is
peaked in the threshold $\tau\approx 1$ region. Even so, the pointlike
determination of the NNLO
contribution to the total hadronic cross section can be off by almost
5-10\%  due to this spurious high energy behaviour, especially for
relatively large values of $\mhiggs\gsim 200$~GeV. Because the NLO and
NNLO corrections to the cross section are quite large, the overall
effect of these terms on the cross section is at the per cent level,
and in particular their effect at NNLO is rather larger than at NLO.

A study of the phenomenological implications of these
results is thus relevant for a precision
determination of the Higgs production cross section.
\smallskip

{\bf Acknowledgements:} We thank Thomas Binoth and Fabio Maltoni for
discussions, and Guido~Altarelli for a critical reading of the
manuscript. This work was partly supported by the Marie Curie
Research and Training network HEPTOOLS under contract
MRTN-CT-2006-035505 and by a PRIN2006 grant (Italy). The work of
S.~Marzani and R.D.~Ball was done with the support of the Scottish
Universities' Physics Alliance.

\appendix
\section{Form factors}
The form factors in eq.~(\ref{Dj}) are given by
\bea
{A}_1(\xi_1,\xi_2,\ytop)&=& C_0(\xi_1,\xi_2,\ytop) \left[\frac{4
    \ytop}{\Delta_3}(1+\xi_1+\xi_2)-1-\frac{4 \xi_1
    \xi_2}{\Delta_3}+12\frac{\xi_1
    \xi_2}{\Delta_3^2}(1+\xi_1+\xi_2)\right] \nonumber \\ 
&-&[B_0(-\xi_2)-B_0(1)] \left[-\frac{2 \xi_2}{\Delta_3}+12\frac{\xi_1
    \xi_2}{\Delta_3^2}(1+\xi_1-\xi_2) \right] \nonumber \\ 
&-&[B_0(-\xi_1)-B_0(1)] \left[-\frac{2 \xi_1}{\Delta_3}+12\frac{\xi_1
    \xi_2}{\Delta_3^2}(1-\xi_1+\xi_2) \right] \nonumber \\ 
&+& \frac{2}{\Delta_3}\frac{1}{(4 \pi)^2}(1+\xi_1+\xi_2) \,,
\eea
\bea
{A}_2(\xi_1,\xi_2,\ytop)&=& C_0(\xi_1,\xi_2,\ytop) \left[2
  \ytop-\frac{1}{2}(1+\xi_1+\xi_2) +\frac{2 \xi_1
    \xi_2}{\Delta_3}\right] \nonumber \\ 
&+&[B_0(-\xi_2)-B_0(1)] \left[\frac{\xi_2}{\Delta_3}(1-\xi_1+\xi_2) \right] \nonumber \\
&+&[B_0(-\xi_1)-B_0(1)]
\left[\frac{\xi_1}{\Delta_3}(1+\xi_1-\xi_2)\right] + \frac{1}{(4
  \pi)^2}\,, 
\eea
with
\beq \label{delta}
%\Delta_3= 1+ \xi_1^2+ \xi_2^2- 2 \xi_1 \xi_2 +2 (\xi_1+\xi_2)\,.
\Delta_3= 1+ \xi_1^2+ \xi_2^2- 2 \xi_1 \xi_2 +2 (\xi_1+\xi_2)=
(1+\xi_1+\xi_2)^2-4 \xi_1 \xi_2\,. 
\eeq
It is also convenient to define the form factor
\beq
\label{athreedef}
A_3(\xi_1,\xi_2,\ytop)\equiv\frac{1}{\xi_1\xi_2}\left[\frac{1+\xi_1+\xi_2}{2}A_1-A_2\right].
\eeq

The scalar integrals $B_0$ and $C_0$ are
\bea \label{b0def}
B_0(\rho)&=& -\frac{1}{8 \pi^2} \sqrt{\frac{4
    \ytop-\rho}{\rho}}\, \tan^{-1}\sqrt{\frac{\rho}{4 \ytop-\rho}}\,,
\quad \textrm{if} \quad 0<\rho< 4 \ytop\,; \nonumber \\ 
B_0(\rho)&=& -\frac{1}{16 \pi^2} \sqrt{\frac{\rho-4 \ytop}{\rho}}
\ln \frac{1+\sqrt{\frac{\rho}{\rho-4
      \ytop}}}{1-\sqrt{\frac{\rho}{\rho-4 \ytop}}}\,,  \quad
\textrm{if} \quad \rho<0 \; \textrm{or}\; \rho> 4 \ytop\,;
\eea
\bea \label{c0def}
C_0(\xi_1,\xi_2)&\equiv& \frac{1}{16 \pi^2} \frac{1}{\sqrt{\Delta_3}}
\Big \{ \ln(1-y_{-}) \ln
\left(\frac{1-y_{-}\delta_1^+}{1-y_{-}\delta_1^-} \right) \nonumber \\ 
&&+\ln(1-x_{-}) \ln
\left(\frac{1-x_{-}\delta_2^+}{1-x_{-}\delta_2^-}\right)+\ln(1-z_{-})
\ln \left(\frac{1-z_{-}\delta_3^+}{1-z_{-}\delta_3^-} \right)  \nonumber \\ 
&& + \textrm{Li}_2(y_+ \delta_1^+)+ \textrm{Li}_2(y_- \delta_1^+)-
\textrm{Li}_2(y_+ \delta_1^-)- \textrm{Li}_2(y_- \delta_1^-) \nonumber \\ 
&& + \textrm{Li}_2(x_+ \delta_2^+)+ \textrm{Li}_2(x_- \delta_2^+)-
\textrm{Li}_2(x_+ \delta_2^-)- \textrm{Li}_2(x_- \delta_2^-) \nonumber \\ 
&& + \textrm{Li}_2(z_+ \delta_3^+)+ \textrm{Li}_2(z_- \delta_3^+)-
\textrm{Li}_2(z_+ \delta_3^-)- \textrm{Li}_2(z_- \delta_3^-)
\quad\Big\}\,, 
\eea
where
\beq \label{deldefin1}
\delta_1 \equiv \frac{-\xi_1+\xi_2-1}{\sqrt{\Delta_3}}\;,\quad
\delta_2 \equiv \frac{\xi_1-\xi_2-1}{\sqrt{\Delta_3}}\;,\quad
\delta_3 \equiv \frac{\xi_1+\xi_2+1}{\sqrt{\Delta_3}}\;,
\eeq
\beq \label{deldefin2}
\delta_i^{\pm}\equiv \frac{1\pm \delta_i}{2}\,,
\eeq
and
\bea \label{xyzdef}
x_{\pm} &\equiv& -\frac{\xi_2}{2 \ytop} \left(1 \pm \sqrt{1+\frac{4
      \ytop}{\xi_2}} \,\right)\,, \nonumber \\ 
y_{\pm} &\equiv& -\frac{\xi_1}{2 \ytop} \left(1 \pm \sqrt{1+\frac{4
      \ytop}{\xi_1}} \,\right)\,, \nonumber \\ 
z_{\pm} &\equiv& \frac{1}{2 \ytop} \left(1 \pm i\, \sqrt{4 \ytop-1}
  \,\right)\,.
\eea 

In the infinite top mass limit the scalar integrals become
\bea
 \lim_{\ytop \to \infty} B_0(\rho) &=& \frac{1}{16 \pi^2} \left(-2 +
   \frac{\rho}{6 \ytop} \right)+ O\left(\frac{1}{\ytop^2} \right) \,,
 \\ 
 \lim_{\ytop \to \infty} C_0(\xi_1,\xi_2) &=& - \frac{1}{32 \pi^2
   \ytop} \left(1 + \frac{1-\xi_1-\xi_2}{12 \ytop} \right)+
 O\left(\frac{1}{\ytop^3} \right)\,, 
\eea
so that the form factors reduce to
\beq
\label{a1a2t}
\lim_{m_t\to\infty} \mtop^2  A_1 = \mhiggs^2  \frac{1}{48 \pi^2};
\qquad\qquad
\lim_{m_t\to\infty} 4 \mtop^2 A_2 = \mhiggs^2 \frac{\as}{48 \pi^2 }\frac{1+\xi_1+\xi_2}{2}.
\eeq
These limits also imply that
\beq
\label{a3t}
\lim_{m_t\to\infty} \mtop^2  A_3 = 0.
\eeq

In the on-shell limit the scalar integrals are
\bea \label{onshellint}
\lim_{\xi_i \to 0} B_0(\xi_i) &=& -\frac{1}{8 \pi^2}\,, \nonumber \\
\lim_{\xi_1 \to 0} C_0(\xi_1,\xi_2,\ytop) &=&
\frac{1}{32\pi^2}\frac{1}{1+\xi_2} \left( \ln^2 \frac{-z_{-}}{z_{+}}-\ln^2 \frac{-x_{-}}{x_{+}} \right) \,,  \\
\lim_{\xi_1, \xi_2 \to 0} C_0(\xi_1,\xi_2,\ytop) &=& \frac{1}{32\pi^2}
\left(\ln^2 \frac{-z_{-}}{z_{+}} \right) \nonumber \,,
\eea
so that
\bea\label{osas}
{A}_1(0,0) &=& \frac{1}{8 \pi^2}  + \frac{1}{32\pi^2} \left(\ln^2 \frac{-z_{-}}{z_{+}} \right)
  \left(4 \ytop-1 \right)\nonumber \\
{A}_2(0,0) &=& \frac{1}{16 \pi^2}  + \frac{1}{32\pi^2} \left(\ln^2 \frac{-z_{-}}{z_{+}} \right)
  \left(2 \ytop-\frac{1}{2} \right)
\eea

The high energy limit of the form factors is trivially determined
when $\xi_1\to\infty$, $\xi_2\to\infty$ with $\xi_1\not=\xi_2$:
\beq\label{hetriv}
\lim_{\xi_1\to\infty,\,\xi_2\to\infty} A_1(\xi_1,\xi_2,\ytop)=0;
\>\lim_{\xi_1\to\infty,\,\xi_2\to\infty}
A_3(\xi_1,\xi_2,\ytop)=0;\>\lim_{\xi_1\to\infty,\,\xi_2\to\infty}
A_2(\xi_1,\xi_2,\ytop)=\frac{1}{(4\pi)^2}.
\eeq
If $\xi_1\to\infty$, $\xi_2\to\infty$ with $\xi_1=\xi_2$ the limit is
more subtle. In this case we get 
\bea \label{a1limit}
 \lim_{\xi \to \infty} A_1(\xi,\xi,\ytop)  &=& \lim_{\xi \to
   \infty}\frac{{\bar C_0(\xi,\xi,\ytop)}}{4} \sqrt{\xi}- \frac{1}{16 \pi^2} \left[\frac{1}{2} \ln \frac{\ytop}{\xi}-1 +
  \sqrt{4 \ytop-1} \tan^{-1} \sqrt{\frac{1}{4 \ytop-1}} \,\right]\nonumber \\ &&+O\left(\frac{1}{\sqrt{\xi}} \right),
\eea
where we have defined
\beq\label{cbardef}
\bar C_0(\xi_1,\xi_2,\ytop)\equiv
C_0(\xi_1,\xi_2,\ytop)\sqrt\Delta_3. 
\eeq
However, it turns out that
\beq\label{czlim}
\lim_{\xi \to \infty} \bar C_0 (\xi,\xi,\ytop) = \frac{1}{16 \pi^2 \sqrt{\xi}} \left[2 \ln\frac{\ytop}{\xi}-4 +4 \sqrt{4 \ytop-1} \tan^{-1} \sqrt{\frac{1}{4 \ytop-1}} \right]
 + O\left(\frac{1}{\xi} \right) \,,
\eeq
hence we conclude that eq.~(\ref{hetriv}) holds also when $\xi_1=\xi_2$.

\end{document}